\documentclass{INTERSPEECH2023}
\usepackage{amssymb}
\usepackage{textcomp}
\usepackage{algorithmic}
\usepackage{algorithm}
\usepackage{multirow, color}
\usepackage{makecell,amsmath}

\interspeechcameraready

\title{TO-Rawnet: Improving RawNet with TCN and Orthogonal Regularization for Fake Audio Detection}
\name{Chenglong Wang$^{1,2}$, Jiangyan Yi$^{2,3}$, Jianhua Tao$^{3,4}$, Chu Yuan Zhang$^{2,3}$, Shuai Zhang$^{4}$, Ruibo Fu$^{2}$, Xun Chen$^{1}$}
\address{
  $^1$University of Science and Technology of China, Hefei, China\\
  $^2$State Key Laboratory of Multimodal Artificial Intelligence Systems, Institute of Automation, Chinese Academy of Sciences, Beijng, China \\
  $^3$School of Artificial Intelligence, University of Chinese Academy of Sciences, China \\
  $^4$Department of Automation, Tsinghua University}
\email{chenglong.wang@nlpr.ia.ac.cn}

\begin{document}

\maketitle
 
\begin{abstract}
Current fake audio detection relies on hand-crafted features, which lose information during extraction. To overcome this, recent studies use direct feature extraction from raw audio signals. For example, RawNet is one of the representative works in end-to-end fake audio detection. However, existing work on RawNet does not optimize the parameters of the Sinc-conv during training, which limited its performance. In this paper, we propose to incorporate orthogonal convolution into RawNet, which reduces the correlation between filters when optimizing the parameters of Sinc-conv, thus improving discriminability. Additionally, we introduce temporal convolutional networks (TCN) to capture long-term dependencies in speech signals. Experiments on the ASVspoof 2019 show that the Our TO-RawNet system can relatively reduce EER by 66.09\% on logical access scenario compared with the RawNet, demonstrating its effectiveness in detecting fake audio attacks.
\end{abstract}
\noindent\textbf{Index Terms}: ASVspoof, fake audio detection, end-to-end, orthogonal convolution

\section{Introduction}

In the field of fake audio detection, the use of standard hand-crafted feature is a common approach \cite{witkowski2017audio, font2017experimental, novoselov2016stc, korshunov2016cross}. Linear frequency cepstrum coefficients (LFCC) feature has been the benchmark feature for various anti-spoofing tasks \cite{todisco2019asvspoof, yamagishi2021asvspoof}. LFCC employs linear filters \cite{sahidullah2015comparison} instead of the traditional Mel filters, which focuses more on the high frequency features than Mel frequency cepstral coefficients (MFCC). The constant Q cepstral coefficients (CQCC) feature is derived from a constant-Q transform (CQT) \cite{brown1991calculation, todisco2017constant}, which is better to capture frequency domain features. In addition, other features such as group delay gram (GD gram) \cite{tom2018end}, log power spectrum (LPS) \cite{das2019long}, and cochlear filter cepstral coefficients instantaneous frequency (CFCCIF) \cite{patel2015combining} have also shown good performance. However, the utilization of standard features may smoothen the speech spectrum, hindering the extraction of vital narrow-band speaker traits, such as pitch \cite{xiao2015spoofing} and formants. On the other hand, direct processing of raw waveforms allows the network to learn low-level embeddings tailored to the specific task.

Recently, more researchers turn their attention to studying methods to directly use raw waveform as system input. \cite{jung2019rawnet, ravanelli2018interpretable, jung2020improved, jung2022pushing}. For instance, Dinkel \cite{dinkel2017end} proposes raw waveform convolutional long short term neural network (CLDNN) to enhance the system's defense against unknown types of attacks. Another study, \cite{ma21d_interspeech} enhances the performance of end-to-end fake audio detection by constructing ResWavegram from the output of one-dimensional convolutions applied to the raw waveform. In addition, a new neural network architecture, SincNet \cite{ravanelli2018interpretable}, has been proposed to improve the feature extraction capability of speech signals by using a set of bandpass filters parameterized with Sinc functions. The adjustable cutoff frequency parameters of SincNet make it perform better in specific tasks compared to traditional Mel filters with fixed parameters. In addition, models such as RawNet2 \cite{jung2020improved} and AASIST \cite{jung2022aasist} also use Sinc functions to directly encode the raw waveform. However, the flexibility of the parameters can also lead to the algorithm getting stuck in suboptimal local minima. Moreover, previous studies have only utilized the structure and initialization parameters of Sinc-conv without iteratively optimizing its parameters \cite{jung2020improved, jung2022aasist, zeinali2019detecting}.

Studies in the field of image processing have shown that orthogonal learned filters can better utilize the model capacity, thereby improving the ability of feature expression and intra-class feature representation \cite{aghdaie2021attention, araujo2021lipschitz, thomas2018learning}. Motivated by \cite{wang2020orthogonal}, we enhance RawNet2 \cite{tak2021end} by introducing an orthogonal constraint. We initialize the convolution kernels in the form of linear-scale Sinc filters using a band-pass filtered Sinc function. By imposing the orthogonality property of matrices to constrain the orthogonal learning process of the convolution kernels, we reduce the correlation between filters. In addition, due to the limitation of convolutional kernel size, CNNs are hard to capture long-term dependencies. Inspired by \cite{bai2018empirical}, this paper uses  temporal convolution network (TCN) instead of Conv1d to expand the receptive field of the convolutional kernel. The proposed method improves the discriminative power of RawNet2 by extracting more robust features from the raw audio signals and capturing the complex temporal dynamics of the speech signals. We demonstrate the effectiveness of our proposed method on two benchmark datasets and show that it outperforms other advanced systems for fake audio detection. Our contributions show that the proposed method can significantly improve the robustness of detection model against spoofing attacks. The main contributions of this study can be summarized as follows:

\begin{itemize}
	\item We proposed a novel deep neural network architecture called TO-RawNet for fake audio detection. The model combines the advantages of orthogonal convolution and TCN to improve upon RawNet2. To our best knowledge, this is the first application of the combination of orthogonal convolution and TCN in the field of fake audio detection.
	\item Compared to RawNet, experiments conducted on the ASVspoof 2019 dataset demonstrate that our TO-RawNet system can significantly reduce EER by 66.09\% in the logical access scenario.
\end{itemize}

\begin{figure*}[t]
	\centering
	
	\includegraphics[height=9.5cm,width=0.98\textwidth]{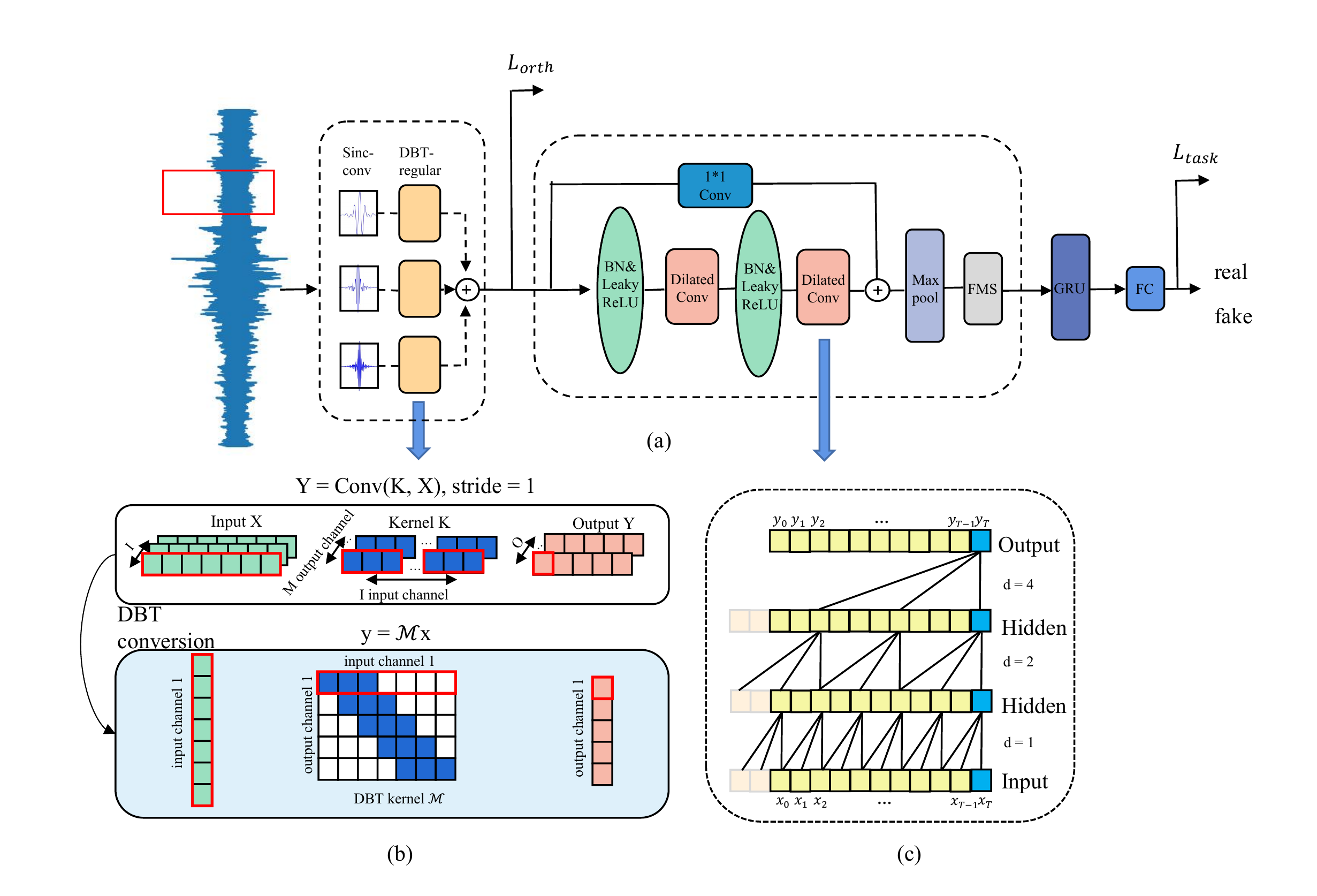}
	\caption{
		(a) Overall framework of the TO-RawNet based fake audio detection model. (b) The convolution expression; $Conv(K, X)$ is converted into a faster DBT vector representation; $y = \mathcal{M}x$. (c) dilated convolution module.
	}
	
	\label{fig:speech_production}
\end{figure*}

The structure of this paper is as follows: Section 2 provides an overview of related work. Section 3 details our proposed method. Experiments, results and discussions are reported in Section 4 and 5, respectively. Finally, we conclude the paper in Section 6.

\section{Related Work}
Recently, more and more researchers have been using raw waveform inputs directly in the field of fake audio detection. SincNet \cite{ravanelli2018interpretable} is a neural network architecture that is designed to operate directly on the raw waveform of audio signals. The first layer of SincNet consists of a bank of band-pass filters that are parametrized as Sinc functions, allowing for the extraction of useful features directly from the raw waveform. By using a constrained first layer with fewer learnable parameters, SincNet is able to learn a more meaningful filterbank structure, resulting in more meaningful output. RawNet2 \cite{tak2021end}, another neural network architecture, also employs a bank of band-pass filters parametrized as Sinc functions to extract features from the raw waveform. The upper layers of RawNet2 consist of residual blocks and gated recurrent units (GRUs) \cite{chung2014empirical}, with the addition of filter-wise feature map scaling (FMS). By applying a sigmoid function to the residual block outputs, FMS serves as an attention mechanism to obtain more distinct representations, resulting in improved discriminative power. Studies have shown that using end-to-end architectures based on learned features rather than knowledge-based and hand-crafted features has the potential to improve the performance of fake audio detection.

\section{Proposed Methods}

\subsection{Orthogonal Convolution}

This paper is based on the differentiable frontend of Sinc-conv, and aims to improve feature expressiveness and intra-class feature representation by using orthogonal convolutions and regularization constraints. The specific operational steps are as follows: as shown in Figure 1 (b), we view the convolution operation as a matrix-vector multiplication, where the kernel matrix $\mathcal{M}$ is generated by the convolution kernel $K$. Using the linear property of the convolution operation, we adopt the Doubly Block-Toeplitz (DBT) matrix construction method to transform the convolution expression $Conv(K,X)$ into a faster DBT matrix-vector representation, as shown below:
\begin{equation}
	Y = Conv(K,X) \Leftrightarrow y = \mathcal{M}x
\end{equation}

where $\mathcal{M}$ is the DBT matrix, and $x$ and $y$ represent the input and output tensors, respectively. The shape of the DBT matrix $\mathcal{M}$ is $K \in R^{(OT_2)\times(IT_1)}$, Where O and I are the output and input channels, and $T_2$ and $T_1$ are the feature map lengths of the output and input. And its rows need to be orthogonalized to reduce the correlation between filters. The orthogonality condition of the DBT matrix $\mathcal{M}$ is shown in equation (2):
\begin{equation}
    \left\langle {\mathcal{M}_{it_1^{\textquotesingle}}, \mathcal{M}_{jt_2^{\textquotesingle}}}\right\rangle = \begin{cases} 1& \text{$(i, t_1^{\textquotesingle}) = (j, t_2^{\textquotesingle})$}\\0& \text{else} \end{cases}	
\end{equation}
where $t_1^{\textquotesingle}$ and $t_2^{\textquotesingle}$ represent two different filter positions, and i and j represent the corresponding row positions of these filters in the matrix. However, since $\mathcal{M}$ is highly structured and sparse, a more efficient method for orthogonal calculation was proposed in \cite{wang2020orthogonal}, as shown in equation (3):
\begin{equation}
	Y = Conv(K, K, padding = P, stride = S) =  I_{r0}
\end{equation}
where $K$ represents the size of the convolution kernel, $S$ represents the stride, and $P=\lfloor {\frac{K-1}{S}} \rfloor \cdot S $ represents padding. $I_{r0}$ is a tensor, where the center is an $n\times n$ identity matrix, and the rest is padded with zeros. By minimizing the difference between $Z = Conv(K, K, padding = P, stride = S)$ and $I_{r0}$, a roughly orthogonal convolution can be obtained. The loss function of the orthogonal convolution can be expressed as:
\begin{equation}
	min_KL_{orth} = {\Vert Z - I_{r0} \Vert}_F^2
\end{equation}
The final training loss is as follows:
\begin{equation}
	L = L_{task} + \lambda L_{orth}
\end{equation}
Where $L_{task}$ represents the loss of the classification task, and $\lambda$ is the weight of the orthogonal regularization loss, and we set three different values in the experiment. Please refer to Algorithm 1 for the specific pseudocode implementation, where $o\_c$ represents output channels and $i\_c$ represents input channels.

\begin{algorithm}[H]
	\caption{Detailed Procedure of Orthogonal Convolution.}\label{alg:alg1}
	\begin{algorithmic}
		\STATE 
		\STATE {\textbf{function} deconv\_orth\_dist(kernel, stride, padding):}
		\STATE \hspace{0.5cm} [o\_c, i\_c, kernel\_size] = kernel.shape
		\STATE \hspace{0.5cm} output = conv\_1d(kernel, kernel, stride, padding)
		\STATE \hspace{0.5cm} target = zeros((o\_c, o\_c, output.shape[-1]))
		\STATE \hspace{0.5cm} center = floor\_divide(output.shape[-1], 2)
		\STATE \hspace{0.5cm} target[:, :, center] = eye(o\_c)
		\STATE \hspace{0.5cm} \textbf{return} norm(subtract(output, target))

	\end{algorithmic}
	\label{alg1}
\end{algorithm}

\subsection{Temporal Convolution Network}
Inspired by TCN \cite{bai2018empirical}, we propose the dilated convolution block to extract features, as shown in Figure 1 (c). The residual block first employs batch normalization and the leaky ReLU activation function, followed by dilated convolution. Next, a 1x1 convolution is used to adjust the output channels to match the input channels. To expedite convergence and facilitate the training of deeper models, we incorporate a residual \cite{he2016deep} path. Each block's output serves as the input for the subsequent block. The dilation factor is doubled for each block up to a certain limit and then repeated ($e.g., 1, 2, 4, ..., 2^n$). This exponential increase in dilation factor ensures that the model captures sufficient temporal contextual information for detecting fake audio. It enlarges the network's receptive field and captures forgery traces in the entire speech with fewer stacked layers.
 
\subsection{TO-Rawnet}
Figure 1 (a) illustrates the architecture of our proposed TO-Rawnet system. First, the raw waveform is fed into the Sinc-conv layer with orthogonal regularization to produce a high-level speech representation. The orthogonal regularization helps reduce redundancy by enabling each filter to focus on distinct frequency components. Next, the high-level representations are fed into the residual module, which includes dilation convolutions with exponentially increasing receptive fields. This enables the network to effectively increase its perception range, allowing it to capture more global information from the input audio. Subsequently, we connect a GRU to extract an utterance-level representation, which is then fed into a softmax activation function to perform real/fake classification.
\section{Experiments}
\subsection{Dataset}
\subsubsection{ASVspoof 2019 Challenge Dataset}
ASVspoof 2019 LA \cite{todisco2019asvspoof} mainly has 19 spoofing attack algorithms (A01-A19), with two types of spoofing attacks: text to speech (TTS) and voice conversion (VC). The LA data set contains three subsets: the training set, the development set, and the evaluation set. Table 1 details the number of real and fake audio of the ASVspoof2019 LA dataset. The attack algorithms in the training and development sets overlap, while the evaluation set includes unseen spoofing attacks.

\subsubsection{ASVspoof 2021 Challenge Dataset}

ASVspoof 2021 LA \cite{yamagishi2021asvspoof} poses greater challenges than the previous versions. Although the training and development sets remain the same as those of ASVspoof 2019 LA database, the evaluation set is distinct. Specifically, the evaluation data for 2021 LA contains encoding and transmission artifacts that stem from actual telephony systems. 

\begin{table}[t]
	
	\caption{ The detailed information of ASVspoof2019 LA dataset and ASVspoof2021 LA dataset.}
	
	\centering
	\begin{tabular}{c||c|c|c }
		\toprule
		\multirow{2}{*}{\makecell[c]{\textbf{Set}}} &\textbf{Genuine}  &  \textbf{Spoofed} &  \textbf{Total} \\
		\cline{2-4}
		& \# utterance & \# utterance & \# utterance \\
		\hline
		Train &2,580   &  22,800  &  25,380  \\
		Dev& 2,548  & 22,296  & 24,844        \\
		Eval(2019 LA) &7,355   & 64,578   & 71,933    \\
		Eval(2021 LA) &18,452 & 163,114 &181,566    \\
		
		\bottomrule
	\end{tabular}
	
\end{table}

\subsection{Experimental Setup}

The audio sampling rate is 16k. To form batches, we standardized the duration of the raw waveform input to approximately 4 seconds (64600 samples) by either truncating longer utterances or concatenating shorter ones. The Sinc-conv layers have a filter length of l = 129, a stride of d = 1, and utilize n = 128 filters. We used fixed linear-scale Sinc filters. To enhance the performance of our model, we utilized six residual-blocks architecture that consists of 12 dilated convolution blocks with varying dilation factors, where the highest dilation factor is 32. In order to prevent over-fitting and under-fitting during the training process, we experimented with different channel combination configurations to determine the optimal combination. The number of channels in the first two residual blocks and the last four residual blocks are set to (32, 64), (128, 256), and (256, 512), respectively. We named them small (S), medium (M), and large (L), in that order. To further improve the discriminative power of our model, we employed FMS independently for each residual-block output. This technique enhances the most informative filter outputs and improves the overall accuracy of the model. To aggregate frame-level representations into an utterance-level representation, we utilized a GRU layer with 1024 hidden nodes. The output of the GRU layer is passed through a softmax activation function, which produces two-class predictions, i.e., real or fake. We propose an orthogonal regularization loss in this paper and set three different weights $\lambda$ for this loss: 0.05, 0.1, and 0.2. 

To train the model, we use the Adam optimizer with a learning rate of $5*10^{-5}$. We set the batch size to 32. The model is trained for 150 epochs. The training set is used to train the model, the development set is used to select the model with the best performance, and finally, the evaluation set is used for evaluation. The results of the ASVspoof2021 competition suggest that data augmentation can reduce overfitting and improve generalization \cite{yamagishi2021asvspoof, tomilov21_asvspoof, tak2022rawboost}. To this end, in our experiments on ASVspoof2021, we utilized data augmentation techniques. Specifically, we employed the open-source tool RawBoost\footnote{https://github.com/TakHemlata/RawBoost-antispoofing} for performing data augmentation in the LA task. We added linear and nonlinear convolutional noise and impulsive signal-dependent additive noise in the LA database. 

In this work, in order to evaluate the results of different fake audio detection systems, the equal error rate (EER) \cite{cheng2004method} is used as the evaluation metric.
\section{Results and Discussion}
\subsection{Ablation Experiments}
Table 2 reveals that the hyperparameters perform better when $\lambda$ is set to 0.1. As a result, for subsequent experiments, we have kept $\lambda$ fixed at 0.1. The Orth-RawNet-M based model consistently outperforms the other models when the same $\lambda$ is used, indicating that the channel combination of (128, 256) prevents over-fitting or under-fitting issues.

Table 3 shows results for ablation experiments for which one of the components in the TO-RawNet model is removed. The results indicate that both orthogonal regularization and TCN have positive effects on the processing of speech signals. Experimental comparisons demonstrate that the TO-RawNet model, which combines these two techniques, performs the best. This is because orthogonal regularization can reduce the correlation between filters, thereby improving the model's generalization ability, while TCN can capture the long-term dependencies in speech signals. Furthermore, all of the models we tested outperformed the baseline system RawNet.

Based on the third column of Table 3, we can draw the conclusion that while the improvement in performance on ASVspoof2021 is less significant than that on ASVspoof2019, the model's performance has still been enhanced through the combination of orthogonal regularization and TCN.

\subsection{Compared with Other Systems}
Compared with the two baseline systems (CQCC-GMM and LFCC-GMM), our method has shown a significant improvement, with EER decreasing from 9.57\% and 8.09\% to 1.58\%, respectively. Compared with neural networks using traditional feature extraction methods \cite{tak2021end, zhang2021one, li2021channel}, the proposed model has better performance. Compared with RawNet2, which also uses raw waveform as input, the proposed method in this paper achieved a relative improvement of 66.09\%. Considering that our method is an improvement upon RawNet2, this demonstrates the effectiveness of our approach. However, compared with the current state-of-the-art (SOTA) single-system AASIST model, the performance of our TO-RawNet model is slightly inferior. The AASIST model also uses raw waveform as input and employs a well-designed graph neural network after Sinc-conv encoding. To verify the effectiveness of our method, we applied orthogonal regularization in the Sinc-conv stage of the AASIST model (since TCN cannot be added to graph neural networks). The results showed that the performance of Orth-AASIST, after orthogonal regularization, was further improved compared to the original AASIST model, with EER decreasing from 1.13\% to 1.02\%. This demonstrates the universal applicability of the orthogonal regularization proposed in this paper.
\begin{table}[!t]
	\caption{Orth-RawNet based models tested on the ASVspoof2019 LA evaluation set. Orth-RawNet-S, Orth-RawNet-M and Orth-RawNet-L with different values of $\lambda$. Results are the average (best) obtained from three runs of each experiment with different random seeds.}
	\centering
	\label{tab:la}
	\begin{tabular}{c||c|c}
		\hline
		\textbf{Methods} & \textbf{$\lambda$} & \textbf{EER} \\
		\hline
		\multirow{3}{*}{\makecell[c]{Orth-RawNet-S}} & 0.05 & 4.39 (4.15)  \\
		& 0.1 &3.82 (3.63) \\
		& 0.2 &4.51 (4.43) \\
		\hline
		\multirow{3}{*}{\makecell[c]{Orth-RawNet-M}} & 0.05 & 3.86 (3.57) \\
		& 0.1 &\textbf{3.19 (3.06)} \\
		& 0.2 &3.52 (3.36) \\
		\hline
		\multirow{3}{*}{\makecell[c]{Orth-RawNet-L}} & 0.05 & 3.78 (3.65)  \\
		& 0.1 & 3.66 (3.59) \\
		& 0.2 &3.94 (3.73) \\
		\hline
	\end{tabular}
\end{table}

\begin{table}[!t]
	\caption{The ablation experiments on the ASVspoof2019 LA and ASVspoof2021 LA evaluation sets are represented by EER1 and EER2, respectively. Results are the average (best) obtained from three runs of each experiment with different random seeds.}
	\centering
	\label{tab:la}
	\begin{tabular}{c||c|c}
		\hline
		\textbf{Methods}  & \textbf{EER1} & \textbf{EER2}\\
		\hline
		RawNet &  4.66 & 5.31 \\
		\hline
		Orth-RawNet-S  & 3.82 (3.63) & 5.02 (4.86) \\
		Orth-RawNet-M  & 3.19 (3.06) & 4.81 (4.73) \\
		Orth-RawNet-L  & 3.66 (3.59) & 4.62 (4.55)\\
		\hline
		TCN-RawNet-S  & 3.43 (3.37) & 5.26 (5.13) \\
		TCN-RawNet-M  & 2.86 (2.62) & 5.08 (4.87) \\
		TCN-RawNet-L  & 3.25 (3.14) & 5.12 (4.96)\\
		\hline
		TO-RawNet-S  & 1.97 (1.86) &  4.05 (3.84)\\
		TO-RawNet-M  & \textbf{1.58 (1.23)} &  \textbf{3.70 (3.58)}\\
		TO-RawNet-L  & 2.56 (2.37) & 3.93 (3.78) \\
		\hline
	\end{tabular}
\end{table}

\begin{table}[!t]
	\caption{Performance comparison of the proposed methods to some known single systems on the ASVspoof2019 LA evaluation set. }
	\centering
	\label{tab:la}
	\begin{tabular}{c||c|c}
		\hline
		\textbf{Methods} & \textbf{Front-end} & \textbf{EER} \\
		\hline
		CQCC-GMM (Baseline1) \cite{nautsch2021asvspoof}  & CQCC &  9.57\\
		LFCC-GMM (Baseline2) \cite{nautsch2021asvspoof} & LFCC &  8.09\\
		\hline
		S1-RawNet2 \cite{tak2021end} &  Raw waveform & 5.64\\
		S2-RawNet2 \cite{tak2021end} &  Raw waveform & 5.13\\
		S3-RawNet2 \cite{tak2021end} & Raw waveform & 4.66\\
		Resnet18-OC-softmax \cite{zhang2021one} & LFCC & 2.19 \\
		MCG-Res2Net50 \cite{li2021channel} & CQT & 1.78 \\
		AASIST \cite{jung2022aasist} & Raw waveform &  1.13\\
		\hline
		TO-RawNet (ours)  & Raw waveform & \textbf{1.58}\\
		Orth-AASIST (ours)  & Raw waveform & \textbf{1.02}\\
		\hline
	\end{tabular}
\end{table}

\section{Conclusions}
We propose a new end-to-end fake speech detection system named TO-RawNet, which has two new contributions: (i) using orthogonal regularization to constrain the learning process of filters, thereby improving the ability of feature expression and intra-class feature representation; (ii) introducing TCN to capture long-term dependencies in time-series data. Compared to RawNet, our TO-RawNet system reduces the EER by 66.09\% in logical access scenarios. Furthermore, we apply the orthogonal regularization technique to the SOTA single-system AASIST and observe performance improvement, verifying the generalizability of orthogonal regularization. In the future, we will verify the performance of TO-RawNet across datasets and further improve its performance on backend models.

\bibliographystyle{IEEEtran}
\bibliography{mybib}

\end{document}